# Threat Characterization: Trajectory Dynamics
(White paper 039)

Russell Schweickart, Clark Chapman, Dan Durda, Piet Hut, Bill Bottke, David Nesvorny[1]

Summary: Given a primary interest in "mitigation of the potential hazard" of near-Earth objects impacting the Earth, the subject of characterization takes on an aspect not normally present when considering asteroids as abstract bodies. Many deflection concepts are interested in the classic geophysical characteristics of asteroids when considering the physical challenge of modifying their orbits in order to cause them to subsequently miss an impact with Earth. Yet for all deflection concepts there are characteristics of the threat which overwhelm these traditional factors. For example, a close gravitational encounter with Earth some years or decades prior to impact can reduce the velocity change necessary for deflection by several orders of magnitude if the deflection precedes the close encounter (or encounters). Conversely this "benefit" comes at a "price"; a corresponding increase in the accuracy of tracking required to determine the probability of impact. Societal issues, both national and international, also characterize the NEO deflection process and these may strongly contend with the purely technical issues normally considered. Therefore critical factors not normally considered must be brought into play as one characterizes the threat of NEO impacts.

## I Introduction

The context for this paper was created in the language of the NASA 2006 Authorization bill, and specifically in the amendment to the Space Act modifying NASA's responsibility with respect to protecting Earth from the potential of near-Earth object (NEO) impacts. The modified language states, *"The Congress declares that the general welfare and security of the United States require that the unique competence of the National Aeronautics and Space Administration be directed to detecting, tracking, cataloguing, and characterizing near-Earth asteroids and comets in order to provide warning and mitigation of the potential hazard of such near-Earth objects to the Earth."*

This context broadens the more usual scientific use of the word characterization as applied to asteroids to focus on those characteristics of NEOs which relate to warning and mitigation of the hazard presented by these objects. It is with this in mind that this paper addresses critical characterization issues not suggested in the "Call for Papers" section on Characterization of NEOs.

Both the challenge of predicting the future path and probability of impact of a threatening NEO and ultimately modifying its orbit in order to cause it to miss impacting the Earth are dramatically shaped, even controlled, by the influence of close gravitational encounters prior to the nominal impact. Therefore the characterization of the orbital dynamics of the NEO of interest, both before and during a deflection, are of great significance when addressing the responsibility now assigned to NASA.

## II Close Encounters, Resonances, & Keyholes

When a NEO experiences a close gravitational encounter with one of the inner solar system planets, Jupiter, or even one of the largest asteroids it experiences a change in its orbit causing subsequent predictions of its orbital path to be altered to some degree. While each of these encounters will affect a NEO and while all of them must be taken into account in tracking NEOs and predicting their impact probabilities, this paper will use only close encounters with Earth (the most common encounter for potentially hazardous asteroids (PHAs)) to illustrate the points made.

Additional note must be made that the detailed computation of orbital parameters and projections into the future entail substantial and complex computation due to the inherent many-body gravitational dynamics involved. Nevertheless, for purposes of illustration many of the most profound effects can be understood using first order simplifications and such will be used in this paper. Additionally, because of the readily available data and illustrations, the two NEOs Apophis (formerly 2004MN4) and 2004VD17 will be used to illustrate most of the points made.

One of the seminal documents in developing an appreciation for the influence of close gravitational encounters in predicting NEO impacts is *Resonant returns to close approaches: Analytical theory*, by Valsecchi, Milani, Gronchi, and Chesley[2]. In this technical paper Valsecchi, et al, introduce the concept of resonant returns and associated keyholes which lie at the heart of many, if not most, NEO impacts with the planets of the inner solar system.

When a NEO of interest[3] makes a close pass by Earth its period is either shortened or extended depending on whether it passes ahead of or behind the Earth as it orbits the Sun. In order to simplify the illustration this paper will assume an encounter passing behind the Earth and therefore an extension of the orbital period. In essence as the asteroid passes behind the Earth it is pulled forward, approximately along its orbital path, thereby increasing its semi-major axis and orbital period. The closer the NEO passes to the Earth the greater is the gravitational impulse and the larger the resultant orbital period.

A gravitational resonance is established if the resultant NEO orbital period is such that, after $n$ orbits the time to pass through the identical orbital position is equal to $x$ years, i.e., after $x$ years the Earth will have orbited the Sun $x$ times while the NEO will have orbited the Sun $n$ times, and both will have returned precisely to their original close encounter positions. The exact distance at which the NEO must pass behind the Earth to establish any specific resonance is unique.

If one holds this image in mind it is clear that if the orbital period of the asteroid is very slightly shorter than the resonance period, it will arrive at the encounter very slightly ahead of schedule. The Earth, therefore, will not yet have arrived at its resonance position and will therefore be closer to the NEO as it crosses the Earth's orbit. Clearly there is a value for the NEO orbital period such that the Earth will be short of its resonance position precisely equal to the original close encounter miss distance, thereby causing an impact.

Since the Earth is an extended object there is, in fact, a small range of NEO periods where, at one extreme the NEO will just graze the trailing edge of the Earth and at the other the leading edge of the Earth. If one works these grazing encounters with Earth's limbs backward they equate to slightly different orbital periods which in turn were caused by slightly different original distances from the Earth at the time of the close encounter.  Since in this example the resultant period of the NEO orbit is slightly less than that required for precise resonance the range of distances from the Earth at the time of close encounter which would result in an impact are slightly greater than the resonance distance.  This range of distances from the Earth at the time of close encounter is called a keyhole.  Each keyhole is associated with a specific resonance value (e.g., 7/6, 8/7, 15/13, etc. i.e. 7 Earth orbits and 6 NEO orbits, etc.) and lies just outside the respective resonance distance.

For example, in the case of Apophis, there is a close encounter with the Earth on April 13, 2029 which sets up the possibility of a 7/6 resonant return impact on April 13, 2036, 7 years after the encounter.  In the case of Apophis the 2029 encounter is unusually close to the Earth with the current projection of its orbit yielding a pass at only 5.94 Earth radii out from the geocenter (i.e. within the geostationary orbit).  This close pass will result in Apophis passing by the Earth on that evening as a magnitude 3 visual object for those in the appropriate longitudinal band to witness it.

The Apophis keyhole, located at approximately 5.73 Earth radii is only 600 meters wide.  If Apophis (or more precisely its center of gravity) passes at the end of that 600 meter region closest to the Earth then in 2036 it will return to just graze the trailing edge of the Earth.  Alternatively if it passes just at the outer edge of the keyhole it will return 7 years later to just graze the leading edge of Earth.  A passage anywhere between those two limits will result in a direct impact with Earth in 2036. (Figure 1)

**III The Deflection Challenge**

The positive aspect of a gravitational keyhole is that unless the NEO passes through the keyhole it will not subsequently impact the Earth at the resonance time.  Therefore the deflection challenge, for asteroids passing through gravitational keyholes prior to impact, is to simply cause them to miss the keyhole.  Since a resonance keyhole is, in general, considerably smaller than the size of the planet itself, a deflection, executed prior to the gravitational close encounter, is a far less demanding deflection than one for which the "target" to be missed is the Earth, per se.

In the case of the 2036 potential Apophis impact, the 2029 keyhole is only 600 meters wide (approximate) and therefore a deflection executed prior to 2029 (should it be required) will require several orders of magnitude less momentum transfer to the asteroid than if the deflection were to be executed subsequent to the close encounter.

This point is dramatically illustrated in the $\Delta V$ plot for Apophis developed by Andrea Carusi[4] (Figure 2).  As can be seen in the figure the minimum $\Delta V$ required to deflect Apophis after the 2029 close encounter is on the order of $2 \times 10^{-2}$ meters/sec, whereas that required prior to 2029 is $10^{-6}$ meters/sec or less.

Another case from the existing database is provided in the instance of 2004VD17.  In this case, however, instead of a single close encounter between the present and the time of potential impact (May 4, 2102) this NEO experiences 3 close encounters with Earth.  Once again, referring to the $\Delta V$ plot (Figure 3) of Andrea Carusi, we see that while the close encounters taken individually are not as dramatic as the 2029 Apophis encounter, together they create a reduction in momentum transfer of approximately 3 orders of magnitude for a deflection executed prior to 2031 compared with one executed in the 2090 time frame.

Variations of this magnitude in the required momentum transfer for a successful deflection are highly significant, and in fact may well determine whether a deflection, with any given technology, can be performed at all. Deflecting a specific asteroid from a potential impact is not then simply a matter of the mass of the asteroid and getting in position for a deflection a decade or so ahead. The deflection challenge will, in many cases, not be a smooth function of time but rather a discontinuous function with one or more unique dates prior to which a deflection is possible and subsequent to which available technology may not be adequate.

**IV The Impact Probability Challenge**

The compensatory "other side of the coin" associated with the advantage of deflection prior to keyhole passage is that exceptional accuracy of the future NEO trajectory is required to know whether a deflection is required. When a close encounter precedes an impact the tracking challenge is to know whether or not the NEO is headed for the associated keyhole, a much smaller target than the Earth itself. In the case of Apophis the keyhole through which the asteroid must pass in order to impact the Earth in 2036 is only 600 meters across. Knowing well ahead of time whether or not the asteroid will pass through this narrow gate is a daunting tracking challenge.

Even in the specific instance of Apophis which has been tracked for well over two years, and for which several radar tracking apparitions have been and will be available, it may require transponder tracking of the asteroid in order to obtain adequately accurate information on its orbit in order to rationally determine whether or not to mount a deflection mission[5]. The transponder needed to provide this increased tracking accuracy would have to be flown to the asteroid with adequate lead time to enable the improved tracking information to be processed and a subsequent deflection executed well in advance of the asteroid keyhole passage.

Unfortunately, to date, the proportion of the potential NEO threat that experiences close gravitational encounters with the Earth (or other body) prior to its nominal impact is unknown.

**V Information Gaps**

One of the most frustrating characteristics of tracking NEO orbits is the discontinuous nature of the data. For relatively small NEOs they are only observable when quite close to the Earth and then only when far enough from the line of sight to the Sun. As a result many NEOs, after discovery, are lost due to the limited initial data and the orbit propagation errors that accumulate prior to the next apparition. Even when the errors are small enough that subsequent apparitions can confidently identify the NEO, there are often many years between successive sighting opportunities.

When a NEO of interest is an Aten (i.e. it spends most of its time inside the Earth's orbit), and especially if its orbital period is close to that of the Earth, there will be episodic opportunities to track the NEO for several years followed by many years with no additional tracking opportunity at all.

Apophis is an example of this Aten characteristic and is about to enter a period of over 6 years with no optical tracking opportunity to improve our knowledge of its orbit. Depending on the specific tracking opportunities with any particular NEO we may well be confronted with a threatening object for which we will have to launch a deflection mission with a relatively low impact probability simply because waiting until the next tracking apparition will provide inadequate time to accomplish a deflection mission.

The clear implication of these orbital tracking gaps is that various mitigation actions, from prioritizing telescope resources, to launching pre-deflection transponders, to launching actual deflection missions, may well have to be undertaken with (or because of) lower quality NEO trajectory accuracy than desired.

## VI Launch windows

Similar phasing considerations also determine the availability of reasonably efficient launch opportunities for missions to NEOs, especially those with large aphelion distances. In many potential NEO deflection cases the most significant energetic challenge will be to deliver a deflection system to a NEO located in an orbit with either a large orbital period, a highly inclined orbit, or both. In general the energy required to get to a NEO is greater than the energy required to deflect it.

It is likely, therefore, that for a comprehensive NEO deflection capability an advanced highly efficient deep space propulsion capability will have to be developed. Such a capability was recently in development as part of NASA's Prometheus Program. Unfortunately this development program was terminated as a result of budget constraints and conflicting priorities.

## VII Deflection decisions

The orbital dynamics of NEOs, both prior to and during deflection, in combination with the residual tracking uncertainties at the time deflection decisions must be made create another issue characterizing the NEO threat; the need for international coordination.

The Path of Risk: The nature of orbital mechanics assures that certain errors in tracking, projected forward in time to a nominal impact many years in the future, will result in large uncertainties of the location of the asteroid along its orbit track. It is this characteristic which leads to very low, but non-zero probabilities of impact with Earth for many newly discovered NEOs. With additional tracking over time this uncertainty generally drops to zero as the error ellipse at the nominal time of impact shrinks and leaves the Earth well outside the bounds of the ellipse. It is the cases where the error ellipse shrinks to several times the size of the Earth but still contains the Earth within it that are of concern re deflection.

On the other hand those orbital elements which define the plane of the NEO orbit and its orientation with respect to the ecliptic firm up very early in the tracking history. As a result, while there is large uncertainty in whether or not an impact will occur, there is surprisingly low uncertainty in the intersection of the NEO orbit plane and the surface of the Earth. In fact after several months of tracking a NEO it is not unusual to see the +/- 3 sigma width of that intersection drop to only several tens of kilometers in width. This intersection then forms a narrow path across the surface of the Earth which changes very slightly with time, that change being primarily a slow decrease of the width of the path.

This narrow corridor (Figure 4), a few kilometers wide and stretching across the face of the planet can legitimately be called the path of risk (PoR). If (with strong emphasis on the "if") a NEO at issue is going to hit the Earth, it will impact somewhere within that narrow corridor. The uncertainty, often very large, is whether, as the error ellipse (referred to by NEO astronomers as the line of variations or LOV) continues to shrink with additional tracking, it will continue to contain the earth within it. But the PoR within which an impact would occur, if the asteroid is indeed on a collision course, does not change until very close to impact when it simply shrinks in length until it is a line across part of the Earth and ultimately a point.

In many, if not most cases however, at the time a deflection must be considered (for timing reasons) the PoR extends fully across the Earth and for many Earth diameters beyond the limbs as well. Therefore the PoR will, in general, cross many national borders and remain constant for decades as a deflection decision is approached. It will not become clear, in many instances, specifically which nation or nations are actually at risk from an actual impact until after a deflection is already underway.

Instantaneous Impact Point: An additional orbital mechanical reality is that the only effective direction in which to push an asteroid in order to deflect it from an impact is parallel to its orbital velocity vector. While changes in velocity cause an ever increasing change in the location of the asteroid along its future orbit, any velocity changes perpendicular to the NEO velocity vector result in small changes in direction which sinusoidally oscillate around the zero condition.

When a deflection maneuver either adds to or subtracts from the NEO's velocity it causes the period of the asteroid to increase or decrease respectively. With respect to a future impact point on Earth slight increase in the period of the asteroid will require slightly more time for it to arrive at the Earth (several orbits later) thereby resulting in the Earth being slightly further ahead in its orbit at the time the NEO arrives. The result of this slight increase in the NEO's velocity then is to impact the Earth slightly toward its trailing limb from the original impact point. Conversely if the deflection maneuver slows down the asteroid slightly it will subsequently hit a bit toward the leading limb of the Earth at impact.

It takes little work to see that these new impact points, and others as the original deflection velocity change grows, lie along the identical path across the Earth as the earlier PoR. In effect the slight velocity changes associated with the deflection maneuver mimic the uncertainties in the tracking which define the LOV.

The implication of this deflection path is that in the process of deflecting an asteroid its instantaneous impact point will migrate, either slowly or in jumps depending on the deflection technique, toward the leading or trailing limbs of the Earth until it passes beyond the Earth entirely.

From the international political point of view, however, in the process of deflecting an asteroid from an impact with Earth at point A, its instantaneous impact point will shift across the Earth, either eastward or westward from A, until it leaves the Earth's surface altogether on completion of the deflection process. During that process however, should the deflection fail, or be inadequate, there will be a new impact point along the original PoR.

Clearly these dynamic trajectory characteristics will assure not only international interest whenever a NEO impact threatens, but in all likelihood a certainty that the international community, hopefully in a coordinated and thoughtful way, will fully participate in all aspects of the deflection decision process.

Based on the above no one serious about protecting the Earth from NEO impacts should imagine that deflection decisions will be made in isolation. Nor will they be made without the value judgements of other people and cultures being strongly expressed and hopefully taken into full consideration.

## VIII Recommendations

Analytic:
1) Statistical determination should be made of the size of the cohort of NEOs with impact probability for which close gravitational encounters will precede nominal impact.
2) Analysis and display of key parameters related to close encounters (keyholes, ΔV plots, etc) should be routinely produced in order to inform mission design and analysis interests for all NEOs with non-zero impact probability (perhaps with some minimal risk threshold)
3) Timing and phasing issues should be given priority attention. There are serious data gaps re NEO tracking and widely space launch window opportunities for many potential deflection cases. These factors need to be made explicit and accessible in order that everyone involved can deal with this reality.
4) NEO Deflection decision-making should, from the outset, assume and account for the participation of the international community.

Operational:
1) NEO tracking assets, in general, are currently being made available as a secondary mission on most of the telescopes being used. Many of the assets are funded by and primarily dedicated to science yet NEO discovery and tracking is not science, it is public safety. This conflict needs to be resolved and the NEO discovery and characterization efforts put on a secure funding basis. Protection of the Earth from NEO impacts should not be placed in a zero-sum budgeting conflict with space science or exploration.
2) Radar tracking is critical to obtaining an adequate and timely knowledge of NEO orbits for rational planning. Yet funding and upgrading, maintenance, and personnel support for the Arecibo Radio Telescope are unreliable at best. This uncertainty should be addressed at the earliest possible time.
3) It is clear from even a cursory look at the challenge presented by NEOs with a high $V_{imp}$ (velocity at impact) that not only will existing launch vehicles be inadequate to deliver mitigation spacecraft to these asteroids, but that chemical propulsion systems in general are inadequate. High performance, deep space propulsion technology must be placed under development and placed on a firm funding basis. Nuclear-electric propulsion (NEP) in particular should be reconsidered since it provides dual capability for both the delivery and execution of NEO deflection systems.

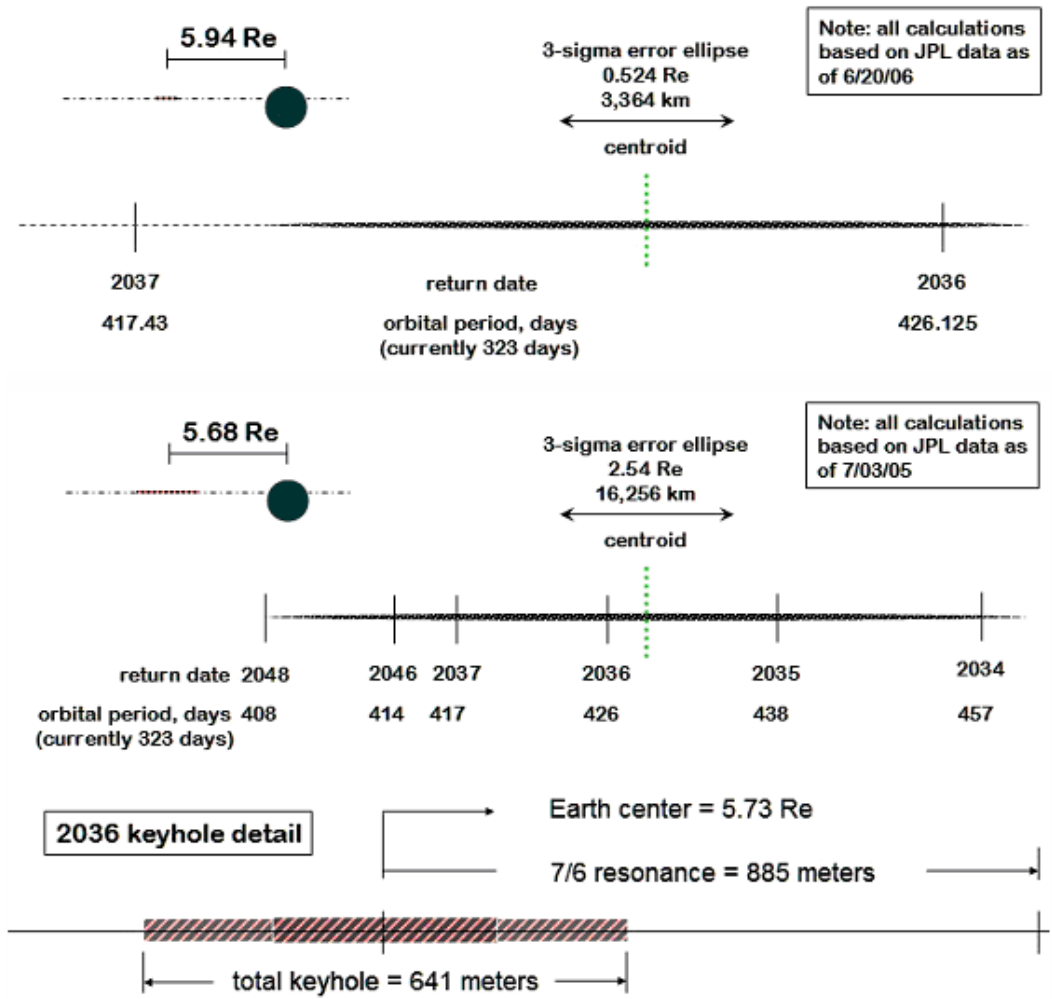

**Figure 1.** Schematic of the Apophis keyhole for various resonant return trajectories. Top, the keyhole geometry for June, 2006 showing the location of the 2036 and 2037 keyholes and the current +/- 3 sigma error ellipse. Middle, the July, 2005 configuration with its much larger error ellipse and many keyholes. Below, detail of the 2036, 7/6 resonance keyhole. (Note: values approximate)

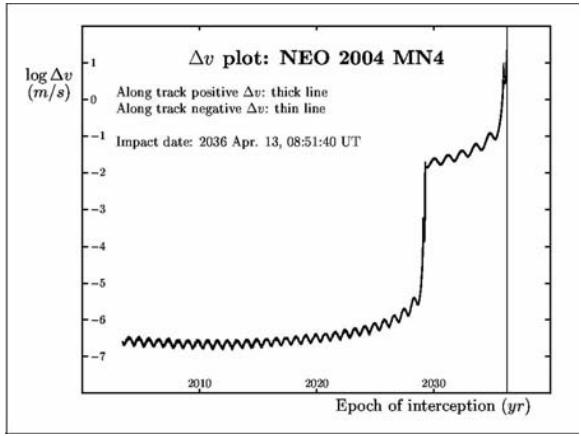 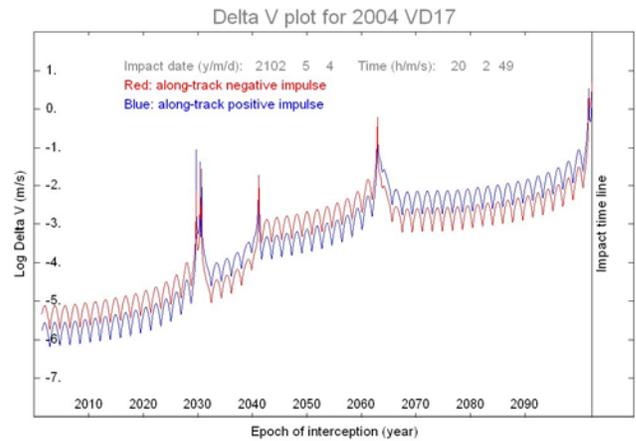

**Figure 2.** ΔV plot for the near-Earth asteroid Apophis (formerly 2004MN4) showing the required deflection change in velocity necessary at various dates prior to the nominal 2036 impact with Earth to cause the asteroid to just miss the impact. The 5 order of magnitude drop in the required ΔV for deflections prior to 2029 is caused by a close gravitational encounter with Earth as Apophis passes within the geostationary satellite orbit on April 13, 2029.

**Figure 3.** ΔV plot for the near-Earth asteroid 2004VD17 showing the deflection change in velocity required to avoid the nominal 2102 impact with Earth. Note that in this case there are three close gravitational encounters prior to the nominal time of impact, each contributing to a 3+ order of magnitude decrease in ΔV for deflections prior to ~2020.

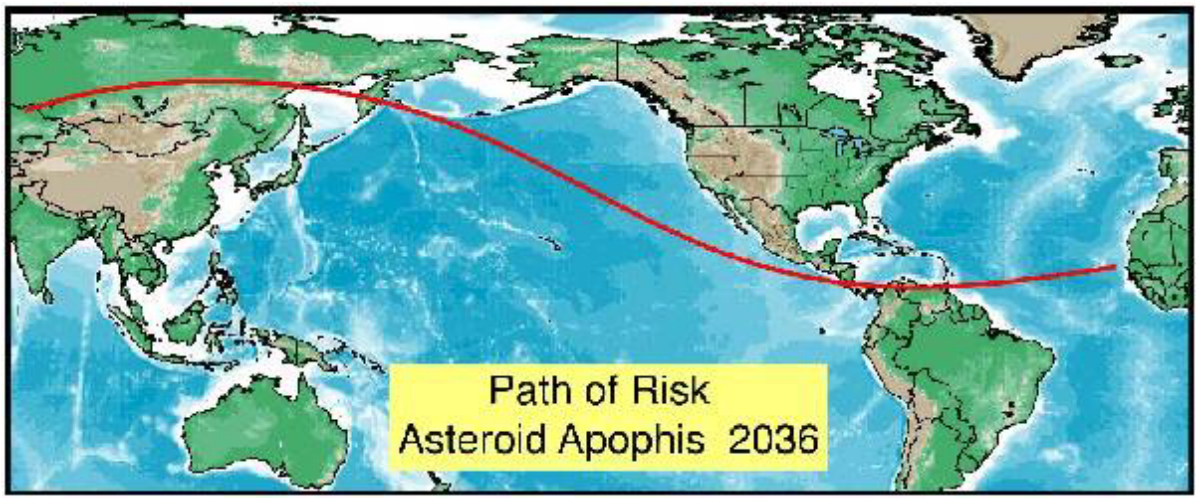

**Figure 4.** Apophis Path of Risk (PoR); the locus of points on the Earth's surface where Apophis could impact, if it were to impact the planet on April 13, 2036. Note that the PoR extends almost 270 degrees around the Earth's surface. The relatively slow speed of the NEO causes gravitational focusing to wrap the potential impact line beyond the Earth's limbs.

[1] Russell L. Schweickart, B612 Foundation; Clark Chapman, Dan Durda, Bill Bottke, David Nesvorny, Southwest Research Institute; Piet Hut, Institute for Advanced Study.

[2] *Resonant returns to close approaches: Analytical theory*, Valsecchi, Milani, Gronchi, and Chesley, Astronomy & Astrophysics, vol. 408, pgs 1179-1196, 2003.

[3] For the purposes of this paper "NEOs of interest" are those large enough to do damage at the Earth's surface and whose orbital path intersects the surface of the Earth as the planet approaches or leaves one or both nodal intersections, i.e., those NEOs which, if the phasing of the asteroid and Earth are precisely aligned, will impact the planet.

[4] Courtesy of the author, Andrea Carusi, University of Torino…

[5] *Potential Impact Detection for Near-Earth Asteroids: The Case of 99942 Apophis (2004 MN)*, Steve Chesley, Asteroids, Comets, Meteors, Proceedings IAU Symposium 229, 2005.